\documentclass[aps,twocolumn,superscriptaddress,showpacs]{revtex4}
\usepackage{amsmath}
\usepackage{amssymb}
\usepackage{amsfonts}
\usepackage[dvips]{graphicx}
\usepackage[]{caption}
\usepackage[]{epsfig}
\begin{document}

\title {Reentrance effect in the lane formation of driven colloids}

\author{J. Chakrabarti}
\affiliation{S. N. Bose National Centre for Basic Sciences, Block-JD,
Sector-III, Salt Lake, Calcutta - 700091, India}
\author{J. Dzubiella}
\email[e-mail address: ] {jd319@cam.ac.uk}
\affiliation{University Chemical Laboratory,
Lensfield Road,
Cambridge CB2 1EW,
United Kingdom}
\author{H. L{\"o}wen}
\affiliation{Institut f{\"u}r Theoretische Physik II,
Heinrich-Heine-Universit\"at D{\"u}sseldorf,
Universit\"atsstra{\ss}e 1, D-40225 D\"usseldorf, Germany}


\date{today}

\begin{abstract}
Recently it has been shown that a strongly interacting colloidal mixture
consisting of oppositely driven particles, undergoes a nonequilibrium
transition
towards lane formation provided the driving strength exceeds a threshold value.
We predict here a reentrance effect in lane formation: for fixed high driving
force and increasing particle densities, there is first a transition towards
lane formation which is followed
by another transition back to a state with no lanes. Our result is obtained
both by  Brownian
dynamics computer simulations and by a phenomenological dynamical density
functional theory. 
\end{abstract}

\maketitle

If a binary colloidal mixture is exposed to a constant external force which
acts differently on the two
particle species, the system achieves a nonequilibrium steady-state in general 
\cite{toprev,PuseyLH,Dhontbook}. For strong external forces,
particles of the same species, which are driven alike, align behind each other
\cite{jo,joprl}. This 
``slip-stream'' effect results
in  macroscopic lanes spanning the whole system size provided a threshold in
the  strength of the
driving force is exceeded. Lane formation can be classified  as a
nonequilibrium first-order phase transition.
The transition towards lane formation is general and occurs
also in systems different from colloids, such as dusty plasma particles
\cite{Konopka}, ions migrating within
two-dimensional membranes \cite{Netz}, granular matter
\cite{valiveti:1999,Herrmann},
 and collective dynamics of pedestrians in pedestrian zones \cite{Helbing3}.

A quantitative investigation of lane formation has been performed by Brownian
dynamics computer simulations
of driven two-component colloidal 
mixture in two and three dimensions \cite{jo}. These simulation studies have
been extended
to asymmetric mixtures \cite{proceedings}, relatively tilted driving fields
acting on the different
particle species \cite{tilted} and crystalline mixtures driven against each
other \cite{tilted}.
Lane formation also occurs as a transient state during the two-dimensional
Rayleigh-Taylor interfacial instability
provided the interfacial tensions are small with respect to the driving force
\cite{Wysocki}.
Moreover a theory for the transition towards lane formation was
recently proposed by us \cite{jay}. It is based on dynamical density functional
theory, 
 which takes the repulsive interparticle interaction explicitly into account,
and is supplemented
by a phenomenological version of the transversal particle current induced by
the external drive. An instability 
from a homogeneous state towards a laned state was predicted
 but the actual predictions for the threshold of the external drive
were significantly smaller than in the simulation~\cite{jo}.

The previous studies in Refs.\ \cite{jo,jay} were restricted
 to high densities of the colloidal particles.
In this brief report, we illustrate our observations on the low density regime 
as well to bring out the entire topology of the laning phase diagram.
We show that there is a {\it reentrant effect} in driven colloidal mixtures.
Reentrance occurs as a function of particle density at fixed strength of the
external drive.
For increasing density and sufficiently high driving force, we find the
sequence:
no lanes/lanes/no lanes, i.e.\ lane formation occurs only in a finite density
window.
There is a simple intuitive reason for this reentrance effect: for very small
densities, the particles are 
almost non-interacting, hence the drive will not induce any instability so that
the system will
stay mixed for entropy reasons. For intermediate densities, there is an
interaction
and the strong drive will put particles into lanes. For very high densities, on
the other hand, 
the interaction is very strong exceeding the external drive so that the
external drive will
again not induce lane formation. Our results are based on Brownian dynamics
computer simulations. We also adopt the dynamical density functional theory
of Ref.\ \cite{jay}. However, we improve the phenomenological expression for
the transverse 
current in this theory and show that the modified theory captures the reentrant
feature and brings about
much better agreement with the simulation data for the location of the laning
transition.

\begin{figure}
  \begin{center}
\includegraphics[width=6.0cm,angle=0.,clip]{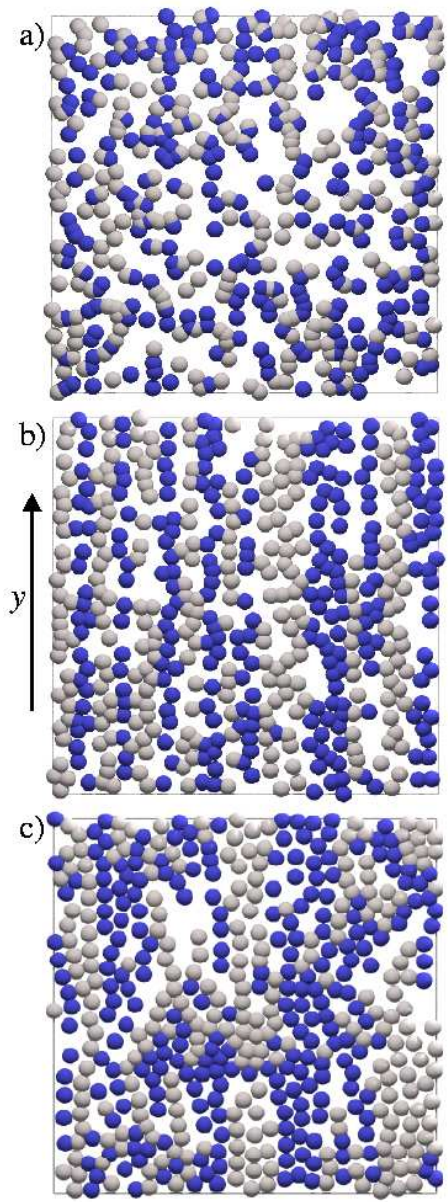}
    \caption{Simulation snapshots for a fixed external force $F^*=70$
    and for densities (a) $\rho_0\sigma^2=0.07$, (b)
    $\rho_0\sigma^2=0.20$, and (c) $\rho_0\sigma^2=0.70$. The driving
    force is acting in $y$-direction as indicated by the arrow. The
    particle and box sizes in this figure are scaled to the same
    lengths for a better comparison of the structure.}
\label{profiles:fig}
\end{center}
\end{figure}

\begin{figure}
  \begin{center}
\includegraphics[width=8.5cm,angle=0.,clip]{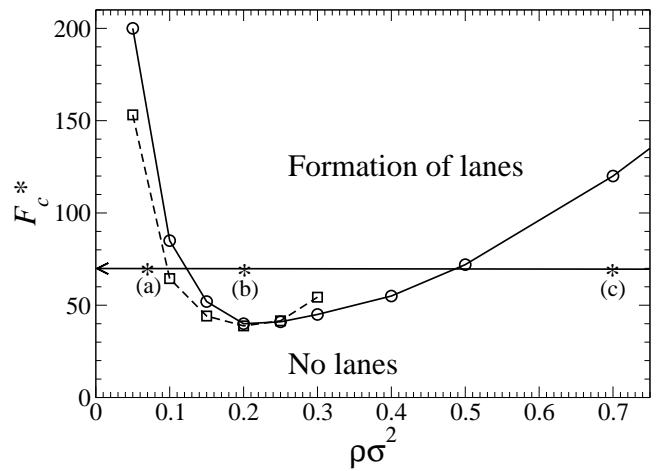}
    \caption{The critical force $F_c$ for lane formation versus
    density $\rho_0$. Brownian dynamics simulation results (circles)
    are compared to the dynamic density functional theory
    (squares). Solid and dashed lines are guide to the eye and suggest
    the phase boundary between states of 'no lanes' and 'lane
    formation'. The asterisks mark states of chosen densities (a)
$\rho_0\sigma^2=0.07$, (b)
    $\rho_0\sigma^2=0.20$, and (c) $\rho_0\sigma^2=0.70$ and a fixed force
$F^*=70$
    (indicated by the arrow), for which snapshots are shown in Fig. 1.}
\end{center}
\end{figure}

The Brownian dynamics simulation model in this work is the same as
used in our previous studies \cite{jo,jay}. Let us briefly summarize
the essential features: we consider an equimolar  binary colloidal system with
$2N=500$ particles in a square cell of length $\ell$, having periodic
boundary conditions with a fixed total number density
$\rho_{0}=2N/\ell^{2}=2\rho^{0}_{+} =2\rho^{0}_{-}$ and a fixed
temperature $T$. 
The effective pair potential between two colloidal
particles at a separation $r$ is modeled as a screened Coulomb
interaction \cite{PuseyLH}
\begin{equation}
V(r) = V_0\,\sigma\exp\left[ -\kappa (r-\sigma)/\sigma\right]/r, 
\label{interaction}
\end{equation}
where $V_0$ is an energy scale, $\sigma$ is the length scale, defining
the range of the interaction, and $\kappa$ the reduced inverse
screening length. In this work the energy is chosen to be
$V_{0}=10k_{B}T$, where $k_{B}T$ is the thermal energy, and
$\kappa=4.0$. The parameters are the same for the interaction between
particles of both types, and also for the cross interaction.  The
dynamics of the colloids is completely overdamped Brownian motion. We
neglect hydrodynamic interactions which is justified for small
colloidal volume fractions and long-ranged colloidal pair
interactions. The constant external field $F_i$ $(i=+,-)$ is acting in
$y$-direction of the simulation cell, shown in Fig.1, and drives particles
of different types in opposite directions: $F_{+}=F>0$ for $(+)$
particles and $F_{-}=-F$ for $(-)$ particles. We denote the
dimensionless drive parameter by $F^{*}=F\sigma/k_{B}T$. In order to
detect lane formation we apply the order parameter $\phi$ which is
defined in \cite{jo}. It is zero, when the system is isotropic, and
one, when the colloids are perfectly ordered in lanes throughout the
whole system, each lane containing only particles of the same type. We
could show in \cite{jo}, that $\phi$, averaged over many steady-state
configurations, increases rapidly to values of about one when critical
threshold value of the applied field, $F_c$ is exceeded, indicating a
nonequilibrium phase transition to the laning state.  We define $F_c$
to be the external field for which $\phi=0.5$, implying that in
average half of the particles in the system are ordered in well
defined lanes. Results for the critical force $F_c$ versus the total
density are plotted in Fig. 2, showing a nonmonotonic behavior: for
fixed high driving force (for instance $F^*=70$ as indicated by an
arrow in Fig. 2) and increasing particle densities, there is first a
transition towards lane formation which is followed by another
transition back to a state with no lanes. Corresponding snapshots for
the three states at three different densities
$\rho_0\sigma^2=0.07,0.20,0.70$ for a fixed force $F^*=70$ are shown
in Fig. 1(a)-(c). The average order parameter for these parameters is
$\phi=0.11,0.80,0.05$, respectively. At the highest density
(Fig. 1(c)), there a still remnant of laning visible but there are also
regions of mutual jamming which destroy perfect lane formation. Hence
there are qualitative differences in the no-lane phase at small and
high densities.

A dynamical density functional theory as described in Ref.\ \cite{jay} accounts
for the stabilization 
of the inhomogeneous steady states beyond the critical drive. The density
fields $\rho_{\pm}(\vec{r},t)$ 
obey the equation of motion which has the form of continuity equation: 
\begin{eqnarray}
\frac {\partial \rho_{\pm}(\vec{r},t)} {\partial t}
 =-\nabla \cdot [ \vec{j}_{\pm}^{(1)}(\vec{r},t) +  
\vec{j}_{\pm}^{(2)}(\vec{r},t) +  \vec{j}_{\pm}^{(3)}(\vec{r},t)].
\end{eqnarray}
The first current term $\vec{j}_{\pm}^{(1)}(\vec{r},t)$ is a diffusive current,
and modeled as \cite{chaikin:lubensky,jay}:
\begin{eqnarray}
\vec{j}_{\pm}^{(1)}(\vec{r},t) = -\frac {D}{k_{B}T} \rho_{\pm}(\vec{r},t)
\nabla
\frac{\delta {\cal F} [\rho_+ , \rho_-   ]}{\delta\rho_{\pm}(\vec{r},t)},
\end{eqnarray}
$D=k_{B}T/\gamma$ being the phenomenological diffusion coefficient. ${\cal
F}[\rho_+,\rho_-]$, the free energy cost 
for creating weak inhomogeneities around a homogeneous state, is given
by \cite{HansenMcDonald}: 
\begin{eqnarray}
\frac{ {\cal F} [\rho_+ , \rho_-]} {k_{B}T} & = & \sum_{k=\pm} \int d^{2}r \rho_{k}(\vec{r}) \log[2\rho_{k}(\vec{r})/\rho_{0}] \nonumber \\
 & - & \frac {1}{2}  \int d^{2}r d^{2}r^{\prime}c(|\vec{r}-\vec{r}^{\prime}|)
\Delta \rho ({\vec{r}}) \Delta \rho ({\vec{r^{\prime}}})
\end{eqnarray}
with $\Delta \rho ({\vec{r}}) =\rho_{+}(\vec{r})+\rho_{-}(\vec{r})-\rho_{0}$
and $c(r)$ the fluid direct correlation 
function, determined by $V(r)$ \cite{HansenMcDonald}.
$\vec{j}_{\pm}^{(2)}(\vec{r},t)$, is induced by the driving 
field in $y$ direction. In a completely demixed state, the field induces a
Brownian drift velocity of 
$v_{d}=F/\gamma$ in the $y$ direction, while in a completely mixed state the
flow is reduced due to mutual collisions between 
oppositely driven particle. These effects can be incorporated via:
\begin{eqnarray} 
\vec{j}_{\pm}^{(2)}(\vec{r},t) & = & \mp {\vec e}_{y} \frac {2v_{d}} {\rho_{0}}
 [1 + \alpha \rho_{\pm}(\vec{r},t)(\rho_{0}/2 \nonumber \\
& - & \rho_{\pm}(\vec{r},t)) 
 -\beta \rho_+(r,t)\rho_-(r,t)],
\end{eqnarray}  
where $\alpha$ is a phenomenological parameter to take care of the appropriate
physical dimensions of the 
current and $\beta$ takes care of the mutual collision effects.
$\vec{j}_{\pm}^{(3)}(\vec{r},t)$ is also induced 
by the external drive but in the $x$ direction transverse to the drive.
Brownian collisions between $+$ and $-$ 
particles having repulsive interaction will cause a displacement perpendicular
to the field. The leading term 
contributing to $\vec{j}_{\pm}^{(3)}(\vec{r},t)$ is the gradient in the density
difference field \cite{jay}. 
$\vec{j}_{\pm}^{(3)}(\vec{r},t)$ should scale with the total collision
cross-section experienced per unit time. 
Let $\sigma_{0}$ denote a typical range of the interaction $V(r)$ needed to
perform a collision and $a_{s}=\rho_{0}^{-1/2}$ 
the mean 
interparticle separation. The number of collision event will be given by the
number of particles of the two 
species in an area of $a_{s}\sigma_{0}$. The typical time of collision is
$a_{s}/v_{d}$, so that the number of 
collision per unit time is
$\frac{v_{d}}{a_{s}}\rho_{+}(\vec{r},t){(0)}\rho_{-}(\vec{r},t)(a_{s}\sigma_{0}
)^2$. 
The cross-section of an individual collision is given by $\pi \sigma_{0}^2/2,$
where the factor of half takes care 
of the fact that the scattering is anisotropic due to the drive, only half
portion of a particle being exposed to 
collision. Consequently we find
\begin{eqnarray}\vec{j}_{\pm}^{(3)}(\vec{r},t) = \pm {\vec e}_{x} \pi \sigma_{0}^2/2
\frac{v_{d}}{a_{s}}\rho_{+}(\vec{r},t){(0)}\rho_{-}(\vec{r},t)(a_{s}\sigma_{0})
^2 \nonumber \\
\times \frac {\partial} {\partial x}  [  \rho_{+}(\vec{r},t) -\rho_{-}(\vec{r},t) ].
\end{eqnarray}
All currents remain unchanged under the interchange of the species, namely,
$\rho_{+}(\vec{r},t) \leftrightarrow 
\rho_{-}(\vec{r},t)$ and $F\rightarrow -F$, which satisfies the symmetry
requirement. $\sigma_{0}$ has been 
estimated by an effective hard core diameter from the Barker-Henderson
perturbation theory, $\sigma_0 = 
\int_0^\infty [1-\exp(-V(r)/k_{B}T)]dr$. 

The dispersion relations for frequency $\omega$ with wave vector ${\vec q} =
(q_{x}, q_{y})$], obtained
from the linearized equations of motion\cite{jay} show that real frequencies,
as required for the steady 
state bifurcations, can result for $q_{y}=0$, independent of $\alpha$ and
$\beta$. The unstable wave vector 
in the $y$ direction being $q_{y}=0$ indicate structural homogeneity in the
$y$-direction parallel to the drive 
as observed in the simulations. Two possible dispersion relations for $q_{y}=0$
are:
\begin{eqnarray}
\omega^{*}_{1} = 2 F^{*} q_{x}^{2} - [1-\rho_{0}{\tilde
c}(q_{x},q_{y}=0)]q_{x}^{2}
\end{eqnarray}
 and  
\begin{eqnarray}
\omega_{2}^{*} = - [1-\rho_{0}{\tilde c}(q_{x},q_{y}=0)] q_{x}^{2},
\end{eqnarray}
where $\omega^{*}_{1,2}=\omega_{1,2}\sigma_{0}^{2}/D$ denotes the
dimensionless frequency and ${\tilde c}({\vec q})$ is the Fourier
transform of $c(r)$. $\omega^{*}_{2}$ remains negative for all
$q_{x}$, since $1-\rho_{0}{\tilde c}({\vec q})$, being the inverse of
the static structure factor, is a positive definite quantity
\cite{HansenMcDonald}.  $\omega^{*}_{1}$ changes sign, indicating the
steady state bifurcation of a homogeneous state to an inhomogeneous
state \cite{prigogine}. $q_{0}$ where $\omega_{1}^{*}$ has a positive
maximum as a function of $q_{x}$, will dominate the growth of the
inhomogeneous phase. We approximate $c(r)$ by that of an effective
hard disk fluid with effective diameter $\sigma_0$ for which
analytical expressions are known \cite{baus}. We calculate numerically
$\omega^{*}_{1}$ and locate graphically the instability point as well
as $q_{0}$ \cite{jay}. For a given $\rho_{0}$ $\omega^{*}_{1}$ remains
negative for all $q_{x}$ indicating the stability of the homogeneous
phase for very low $F^{*}$. The homogeneous phase becomes unstable for
$F^{*}>F_{\rm c}$ to density perturbations with $q_{x}$ over a band
about the maximum at $q_{x}=q_{0}$. Note that a state with density
modes having $q_{y}=0$ but a finite $q_{x}=q_{0}$ indicates a lane
phase as observed in the steady states of the BD simulations. The
stability of the different phases has been shown over the
$F^{*}$-vs-$\rho_{0}$ plane in Fig. 2, $F_{\rm c}^{*}$ being marked by
the squares for chosen densities..  The most striking feature of the
stability diagram is the reentrant homogeneous phase as one increases
the density for a fixed external force.  Physically, the diffusive
thermodynamic current tries to avoid any $+/-$ interface via entropy
of mixing, while the current $\vec{j}_{\pm}^{(3)}(\vec{r},t)$
amplifies particle segregation in $x$ direction via collisions induced
by the drive. The diffusion becomes slower with increasing $\rho_{0}$
that reduces the tendency of mixing. This requires stronger drive to
ensure the collision between two different species to have an
appreciable $\vec{j}_{\pm}^{(3)}(\vec{r},t)$ contribution. This is
what seen in the large $\rho_{0}$ regime where $F_{\rm c}$ increases
with $\rho_{0}$.  However, the collision frequency itself goes down in
the low $\rho_{0}$ regime, and one needs stronger drive to
counterbalance the fast diffusing current, leading to the reentrant
behavior seen in Fig. 2.

In conclusion, we predict a reentrance effect in binary mixtures of oppositely
driven colloids
for increasing density. At fixed driving strength, there is a state with no
lanes
at small densities. This transforms into a state with lanes for intermediate
densities
and then there is reentrance of the state with no lanes. The reentrance
behavior has a simple
intuitive interpretation and 
should therefore be robust with respect to a variation of details of the
interaction.
We expect that it will show up also for asymmetric mixtures, in three spatial
dimensions,
and for different particle dynamics than the simple Brownian dynamics assumed
in our model.
In fact, recent simulations \cite{Padding} which include hydrodynamic
interactions 
have revealed that the laning transition itself is indeed stable when explicit
hydrodynamic interactions
are added. 

The reentrance behavior should be verifiable in experiments. We think that the
most promising
candidate for an experimental test are binary mixtures of 
highly charged colloidal suspensions in a gravitational field. Recent progress
has been made which
highly deionized suspensions \cite{Paddy} which can be subject to sedimentation
at different colloidal densities.
Another possibility are mixtures of colloids and polymers under gravity
\cite{Aarts}. 
The reentrance effect may finally provide also a quantitative framework
to explain different states of pedestrian dynamics in pedestrian zones for
varied  population density \cite{Helbingneu}:
The low-density state without lanes is  typical for low populations. For
intermediate densities,
pedestrians form lanes, but for even higher densities the lanes are broken
again and there is a 
 crossover to panic dynamics.

\acknowledgments We thank H. K. Janssen for helpful discussions and
the DFG for support within the SFB TR6. J.D. acknowledges the
financial support from the EPSRC within the Portfolio Grant RG37352.

\end{document}